\documentclass[12pt,twoside]{article}

\usepackage{amsmath}
\usepackage{amssymb}
\usepackage{epsfig}
\usepackage{graphicx,color}
\usepackage[sort&compress]{natbib}

\bibpunct{[}{]}{,}{n}{}{,}

\newcommand{\bra}[1]{\ensuremath{\langle #1 |}}   
\newcommand{\ket}[1]{\ensuremath{| #1 \rangle}}   
\newcommand{\sprod}[2]{\ensuremath{\left\langle #1 |%
                     #2 \right\rangle}}  

\newcommand{\ltol}{{\ensuremath{\nu_l \rightarrow \nu_l}}}

\newcommand{\MB}{M\"{o}ssbauer}

\title{Comment on `Time-energy uncertainty relations for neutrino oscillations and the
       M\"{o}ssbauer neutrino experiment'}
\author{Evgeny Kh.~Akhmedov%
        \footnote{\protect\parbox[t]{\textwidth}{Also at:
        National Research Centre Kurchatov Institute,
        123182 Moscow, Russia \newline Email: akhmedov@mpi-hd.mpg.de}} ,
        Joachim Kopp\footnote{Email: jkopp@mpi-hd.mpg.de} ,
        Manfred Lindner\footnote{Email: lindner@mpi-hd.mpg.de} \\[0.2cm]
         {\it Max--Planck--Institut f\"{u}r Kernphysik}, \\
         {\it Postfach 10 39 80, 69029 Heidelberg, Germany}}
\date{October 3, 2008}

\makeatletter
\def\@maketitle{%
     \renewcommand{\thefootnote}{\alph{footnote}}
     \newpage
     \vspace*{0.5em}
     \begin{center}%
     \let \footnote \thanks
       {\Large\bf \@title \par}%
       \vskip 1.0em%
       {\normalsize
         \lineskip .5em%
         \begin{tabular}[t]{c}%
           \@author
         \end{tabular}\par}%
       \vskip 0.7em%
       {\normalsize \@date}%
     \end{center}%
     \par
     \vskip 0.5em}
\makeatother

\makeatletter
\renewcommand\section{\@startsection {section}{1}{\z@}%
                                   {-3.5ex \@plus -1ex \@minus -.2ex}%
                                   {2.3ex \@plus.2ex}%
                                   {\normalfont\large\bfseries}}
\makeatother

\begin{document}

\maketitle

\begin{abstract}
  We discuss the implications of the time-energy uncertainty relation to
  recoillessly emitted and captured neutrinos (\MB\ neutrinos) and show 
  that it does not preclude oscillations of these neutrinos, contrary to 
  a recent claim (J.~Phys.~G35 (2008) 095003, arXiv:0803.0527). 
\end{abstract}

\section{The time-energy uncertainty relation for \MB\ neutrinos}

In a recent interesting article~\cite{Bilenky:2008ez}, Bilenky
et al.~considered implications of the time-energy uncertainty relation to
neutrino oscillations. The authors applied their general results to
recoillessly emitted and captured neutrinos 
(\MB\ neutrinos,~\cite{Raghavan:2005gn,Raghavan:2006xf}) and concluded that 
oscillations of such neutrinos would be in conflict with the time-energy 
uncertainty relation. They also suggested that a \MB\ neutrino oscillation 
experiment could test whether the time-energy uncertainty relation is 
applicable to \MB\ neutrinos. 

We believe that the time-energy uncertainty relation, being based on 
fundamental principles of quantum theory, does apply to \MB\ neutrinos.  
Therefore the conclusions of~\cite{Bilenky:2008ez} are in conflict 
with the results of our recent detailed quantum field theoretical 
calculation~\cite{Akhmedov:2008jn}, in which, with no a priori assumptions 
on the propagating neutrino and very well-established assumptions on the 
properties of the source and the detector, we have shown that 
oscillations of \MB\ neutrinos do occur. We will show now that the 
contradiction between the conclusions of \cite{Bilenky:2008ez} and 
\cite{Akhmedov:2008jn} is due to an incorrect application of the
general results of \cite{Bilenky:2008ez} to \MB\ neutrinos.  

The argument in~\cite{Bilenky:2008ez} is based on the Mandelstam-Tamm relation 
\begin{align}
  \Delta E  \Delta O \,\geq\, \frac{1}{2} \Big| \frac{d}{dt} 
\overline{O}(t) 
  \Big|\,,
  \label{eq:MT}
\end{align}
where $O$ is an arbitrary quantum mechanical operator in the Heisenberg
representation, and $\overline{O}(t) = \bra{\psi} O \ket{\psi}$ is its
expectation value in a state $\ket{\psi}$. Choosing $O$
to be the projection operator onto the neutrino flavour $\ket{\nu_l}$,
i.e.~$O \equiv \ket{\nu_l}\bra{\nu_l}$, one can derive the uncertainty 
relation
\begin{align}
  \Delta E\, \geq\, \frac{1}{2} \frac{| \frac{d}{dt} P_\ltol(x,t) |}
                               {\sqrt{P_\ltol(x,t) - P_\ltol^2(x,t)}}\,.
  \label{eq:DeltaE}
\end{align}
Here $P_\ltol(x,t) = |\sprod{\nu_l}{\Psi(x,t)}|^2$, with $\Psi(x,t)$ being
the neutrino wave function, is the probability for finding a neutrino of
flavour $l$ at position $x$ and time $t$.

The authors of~\cite{Bilenky:2008ez} have written $P_\ltol$ in Eq.~(33) of
their paper, which is their version of Eq.~(\ref{eq:DeltaE}), as a function of
only time. They do so because they seek to formulate their arguments not within
quantum mechanics (QM), but within the more general framework of quantum field
theory (QFT), where one often deals with $x$-independent asymptotic states.  It
is indeed possible to define a coordinate independent quantity $P(t) =
|\sprod{\nu_l}{\psi(t)}|^2$ by interpreting $\ket{\psi(t)}$ not as a wave
function, but as a quantum field theoretical state.  However, such an
$x$-independent quantity $P(t)$  has no physical meaning; in particular, it
cannot be interpreted as an oscillation or survival probability unless the
assumption
\begin{equation}
  x \simeq t \,
  \label{eq:xt}
\end{equation}
(``space-to-time conversion'') is invoked. This seemingly innocent assumption,
which is often made for relativistic neutrinos from conventional sources, is
grossly invalid for \MB\ neutrinos. Indeed, it is, strictly speaking, only
correct for {\sl pointlike} relativistic neutrinos or, more generally, in the
case when the size of the neutrino wave packet is small compared to the
distance $x$ traveled by neutrinos. This is not the case for \MB\ neutrino
experiments, for which the baselines of interest are of order of tens to
hundreds of meters, whereas the lengths of the neutrino wave packets exceed
10~km because of near monochromaticity of \MB\ neutrinos ($\Delta E \lesssim
10^{-11}$ eV \cite{Potzel:2006ad}).

In~\cite{Bilenky:2008ez}, the authors obtain their main result by integrating
their Eq.~(33) (the coordinate independent version of our Eq.~\eqref{eq:DeltaE})
over time. Apart from the lack of physical meaning of the integrand, also
the choice of the integration interval is problematic. In~\cite{Bilenky:2008ez},
the integral runs from $0$ to $t_{1min}$, where $t_{\rm 1min} \equiv 2\pi E /\Delta
m^2$ is supposed to be the time it takes the neutrino to travel to the first
oscillation maximum (i.e.~to the first minimum of the survival probability).
Here $\Delta m^2$ is the neutrino mass squared difference, and
following~\cite{Bilenky:2008ez} we have adopted the two-flavour approximation
for neutrino oscillations. However, from the fact that the size of the neutrino
wave packet is much larger than the baseline it is clear that the arrival time
is not well defined for \MB\ neutrinos. It therefore makes no sense to
integrate Eq.~(\ref{eq:DeltaE}) over time, because the integration interval
cannot be given a clear physical meaning.  Instead, we will proceed by
considering the unintegrated version of the time-energy uncertainty relation,
i.e.~Eq.~\eqref{eq:DeltaE} itself.%
\footnote{Note that the $P_\ltol\,$-dependent ratio on the right hand side
of Eq.~(\ref{eq:DeltaE}) can be considered as the reciprocal of the effective
time scale $\Delta t$ over which the expectation value of $O \equiv 
\ket{\nu_l}\bra{\nu_l}$ in the state $\Psi(x,t)$ varies significantly, so
that Eq.~(\ref{eq:DeltaE}) is equivalent to $\Delta E \Delta t \geq 1/2$.}

Following~\cite{Giunti:1997wq}, we write the oscillation probability 
$\nu_l \to \nu_l$ (i.e.~the survival probability of the flavour eigenstate 
$\nu_l$) as
\begin{align}
  P_\ltol(x,t) = \sum_{j,k} |U_{lj}|^2 |U_{lk}|^2 \,
                 e^{- 2 i \phi(x, t)} \,
                 g(x - v_j t) g(x - v_k t)^*\,,
  \label{eq:Pll}
\end{align}
Here $g(x - v_j t)$ are the wave packet shape factors which depend on the 
group velocities $v_j$ of the mass eigenstates and on the width and 
shape of the neutrino wave packets, and $\phi$ is the oscillation phase, 
given by
\begin{equation}
   2 \phi(x, t) = (E_j - E_k) t - (p_j - p_k) x \,.
\label{eq:phi}
\end{equation}
Eq.~(\ref{eq:Pll}) is valid in the limit of no wave packet spreading, which
is a very good approximation for neutrinos. The shape factors allow one to
describe possible effects on oscillations of decoherence and of lack of
localization of the neutrino emitter and absorber. As has been shown
in~\cite{Akhmedov:2008jn}, the coherence and localization conditions
should be very well fulfilled in any realistic \MB\ neutrino experiment,
so $g(x - v_j t)$ can be set equal to unity in the following.
The probability $P_\ltol(x,t)$ then takes the standard form 
\begin{equation}
P_\ltol(x,t) = 1 - \sin^2 2\theta\sin^2 \phi(x, t)\,,
\label{eq:Pll2}
\end{equation}
where $\theta$ is the two-flavour mixing angle. 
Substituting it into Eq.~\eqref{eq:DeltaE}, one readily finds 
\begin{align}
  \Delta E \,\geq \, |E_1 - E_2| \,
                \frac{\sin 2\theta \, \cos\phi(x,t)}
                     {\sqrt{1 - \sin^2 2\theta \sin^2 \phi(x,t)}}\,.
  \label{eq:DeltaE-2}
\end{align}
It is sufficient to consider the case $\sin^2 2\theta = 1$, because
the right hand side has a maximum as a function of $\theta$ then. Phrased 
differently, \eqref{eq:DeltaE-2} is certainly fulfilled if it is fulfilled 
for $\sin^2 2\theta = 1$. In this case the inequality \eqref{eq:DeltaE-2}  
amounts to
\begin{align}
\Delta E \,\geq \, |E_1 - E_2|\,.
\label{eq:DeltaE-3}
\end{align}
Eq.~(\ref{eq:DeltaE-3}) expresses the obvious requirement that the energy 
uncertainty of the neutrino state be larger than the difference of the 
energies of different mass eigenstates composing the given flavour state 
$\nu_l$. It has to be fulfilled in any oscillation experiment, and will 
certainly be satisfied in \MB\ neutrino experiments where, 
due to the large momentum uncertainty of the emitted neutrino state, 
the energy difference $|E_1 - E_2|$ can be vanishingly small without 
violating the energy-momentum relation of relativistic neutrinos 
\cite{Akhmedov:2008jn}.

\section{Evolution in time vs.~evolution in space and time}

In the literature, there exist different approaches (or schemes) for describing 
neutrino oscillations (``oscillations in time'', ``oscillations in space'', 
``oscillations in space and time''). The authors of~\cite{Bilenky:2008ez} 
assert that only the experiment can decide which scheme is the correct one, 
and argue that in fact only the \MB\ neutrino experiments can do the job.  
While we do not consider the theory of neutrino oscillations to be finished or 
closed, we believe that the standard QFT cannot yield different predictions 
for the same process. In our opinion, there exist different approximations 
(not mechanisms or schemes), and to find out which of them are justified, one 
does not need to perform an experiment: it is sufficient to carefully examine 
the validity of the invoked assumptions in each particular case.

In~\cite{Bilenky:2008ez}, the approximation of ``oscillations in time'' is
advocated. We consider this approximation to be invalid for \MB\ neutrinos for
several reasons. Firstly, it is not possible to define a ``time of flight'' for
these neutrinos because they are produced and absorbed recoillessly, and with no
accompanying charged leptons being emitted from the atom. Thus, detection
of the nuclear recoil or of accompanying charged leptons cannot be used for
a precise determination of the neutrino emission or absorption time. It is easy
to see that a detection of the recoil of the crystal as a whole cannot be used
for this purpose either. Indeed, it would require very long times because one
would have to detect a microscopic momentum transfer to a macroscopic body (the
time necessary for the crystal to be displaced by an interatomic distance is
$\gtrsim 10^{10}$~s).  Taking into account also the fact that the length of
the \MB\ neutrino wave packets is about 10~km, we see that the uncertainty
of the neutrino emission and absorption times greatly exceeds the time it
would take a classical relativistic pointlike particle to travel from 
the source to the detector.

As another argument against the ``evolution in time'' picture, note that, if
this picture were true at a fundamental level, i.e. without any ``space-to-time 
conversion'', no far detectors would be required in oscillation experiments 
because the oscillation probability would depend only on $t$, not on $x$. It 
is only through the assumption $x \simeq t$, Eq.~\eqref{eq:xt}, that the 
standard oscillation phenomenology is recovered. However, as we have shown 
above, Eq.~\eqref{eq:xt} does not hold for \MB\ neutrinos. 

One possible argument for ``evolution only in time'' is that such a description
is usually employed (and is known to work well) for oscillations of neutral $K$
and $B$ mesons.  This actually corresponds to going into the rest frame of the
mesons and considering their evolution with proper time.  While this approach
is justified for $K$ and $B$ mesons, which are extremely degenerate in mass, it
is not necessarily applicable to neutrinos, for which the rest frame of flavour
states may simply not exist. Indeed, if neutrino masses are hierarchical, in
the reference frame where one of the mass eigenstates composing a given flavour
state is at rest, the others will be relativistic.  More importantly, neutral
$K$ and $B$ mesons are not even nearly as monochromatic as \MB\ neutrinos, so
that their wave packets are of microscopic size, and for all practical purposes
they can be considered pointlike. As we discussed above, this is not the case
for \MB\ neutrinos, for which the coordinate dependence cannot be ignored even
in their rest frame (if it exists), simply because their wave packets are of
macroscopic size.

The authors of~\cite{Bilenky:2008ez} have correctly pointed out that in QFT the
evolution of the states is described by the Schr\"odinger equation. This does
not, however, mean that the evolution in QFT occurs only in time: In fact, the
Schr\"odinger equation of QFT results in just the standard Feynman rules, which
can also be obtained from the covariant Lagrangian formalism and which describe
the space-time development of the processes.  Since we have used the standard
Feynman rules in our calculations in~\cite{Akhmedov:2008jn}, the approach based
on using the Schr\"odinger equation must yield results identical to ours. 
One may then wonder why the approach of Bilenky et al.~actually gives different
results. In our opinion, this is related to their complete disregard of the
spatial evolution of flavour-eigenstate neutrinos.  Unlike in QM, in QFT the
production and detection processes for mixed states have to be included into
the consideration. This brings in the necessary dependence of the transition
probability on the coordinate (through the coordinates of the neutrino source
and detector). Moreover, this ensures that the asymptotic (i.e.~$in$- and $out$-)
states are mass eigenstates, as they have to be in the standard QFT.

Let us finally comment briefly on the possibly counterintuitive result, used in
several places throughout this article, that \MB\ neutrino wave packets have
macroscopic spatial and temporal extents $\sigma_x \simeq \sigma_t \sim 10$~km.
In a QM approach, this follows immediately from the time-energy uncertainty
relation applied to the production process, which tells us that $\sigma_x
\simeq \sigma_t \sim 1/\Delta E$, where $\Delta E$ is the energy uncertainty
associated with the emission process.  This relation was confirmed
in~\cite{Akhmedov:2008jn} by direct calculations performed within QFT. For all
the regimes we have considered there (inhomogeneous broadening as well as
homogeneous broadening, including the case of the natural linewidth dominance)
we invariably found that the coherence length for \MB\ neutrinos was given by
$L_{\rm coh}\sim 1/\Delta E \, \Delta v_g$, with $\Delta E$ the corresponding
neutrino linewidth and $\Delta v_g$ the difference of the group velocities of
the wave packets corresponding to different mass eigenstates.  Comparing this
to the standard expression%
\footnote{This expression can be easily understood if one notes that 
decoherence occurs after the wave packets corresponding to different 
mass eigenstates have separated in coordinate space. This happens
after a distance $\sigma_x / \Delta v_g$.} 
$L_{\rm coh}\sim \sigma_x/\Delta v_g$, we find $\sigma_x \sim 1/\Delta E$.
Thus, our conclusion that the lengths of \MB\ neutrino wavepackets greatly
exceed the source--detector distance holds both within QM and QFT.

\section{Conclusions}

We conclude that, while the general results of Bilenky et
al.~\cite{Bilenky:2008ez} on implications of the time-energy uncertainty
relation to neutrino oscillations are mostly correct, their application of
these results to \MB\ neutrinos was flawed.  A proper interpretation of the
time-energy uncertainty relation is fully consistent with oscillations of \MB\
neutrinos. 

The main reason for the incorrect conclusion of Bilenky et al.~regarding \MB\
neutrinos was their improper treatment of the evolution of the neutrino state
in QFT.   In our opinion, the only meaningful way to treat neutrino
oscillations in QFT is to explicitly include the production and detection
processes, so that the neutrino appears only as an intermediate state. Since
the source and the detector are spatially localized, the oscillation
probability in this approach exhibits the proper coordinate dependence, i.e.~it
describes the evolution in space and time. The ``evolution in time''
approximation could only be justified by using the assumption $x \simeq t$,
which implies the equivalence of ``evolution in space'' and ``evolution in
time''. Such an equivalence indeed holds for neutrinos from conventional
sources, but not for \MB\ neutrinos, for which the distance traveled is well
defined by that between the source and the detector, while evolution in time
has no clear physical meaning because of the very large lengths of the neutrino
wave packets.

\vspace*{1.5mm} 
We are grateful to S. Bilenky, F. von Feilitzsch and W. Potzel for many useful 
discussions clarifying their point of view. 

\vspace*{1.5mm} 
{\em Note added}. After the first version of this comment (arXiv:0803.1424v1) 
was submitted to the archive, the paper \cite{Bilenky:2008dk} has appeared, in 
which the authors rejected our criticism. In the present version of our paper 
we both comment on~\cite{Bilenky:2008ez} and answer the criticism presented 
in \cite{Bilenky:2008dk}.

\begin{center}
  \rule{10cm}{0.25pt}
\end{center}
\vspace{-1.5cm}


\begin{thebibliography}{7}
\expandafter\ifx\csname natexlab\endcsname\relax\def\natexlab#1{#1}\fi
\expandafter\ifx\csname bibnamefont\endcsname\relax
  \def\bibnamefont#1{#1}\fi
\expandafter\ifx\csname bibfnamefont\endcsname\relax
  \def\bibfnamefont#1{#1}\fi
\expandafter\ifx\csname citenamefont\endcsname\relax
  \def\citenamefont#1{#1}\fi
\expandafter\ifx\csname url\endcsname\relax
  \def\url#1{\texttt{#1}}\fi
\expandafter\ifx\csname urlprefix\endcsname\relax\def\urlprefix{URL }\fi
\providecommand{\bibinfo}[2]{#2}
\providecommand{\eprint}[2][]{\url{#2}}

\bibitem[{\citenamefont{Bilenky
  et~al.}(2008{\natexlab{a}})\citenamefont{Bilenky, von Feilitzsch, and
  Potzel}}]{Bilenky:2008ez}
\bibinfo{author}{\bibfnamefont{S.~M.} \bibnamefont{Bilenky}},
  \bibinfo{author}{\bibfnamefont{F.}~\bibnamefont{von Feilitzsch}},
  \bibnamefont{and} \bibinfo{author}{\bibfnamefont{W.}~\bibnamefont{Potzel}}
  (\bibinfo{year}{2008}{\natexlab{a}}), \eprint{arXiv:0803.0527 [hep-ph]}.

\bibitem[{\citenamefont{Raghavan}(2005)}]{Raghavan:2005gn}
\bibinfo{author}{\bibfnamefont{R.~S.} \bibnamefont{Raghavan}}
  (\bibinfo{year}{2005}), \eprint{hep-ph/0511191}.

\bibitem[{\citenamefont{Raghavan}(2006)}]{Raghavan:2006xf}
\bibinfo{author}{\bibfnamefont{R.~S.} \bibnamefont{Raghavan}}
  (\bibinfo{year}{2006}), \eprint{hep-ph/0601079}.

\bibitem[{\citenamefont{Akhmedov et~al.}(2008)\citenamefont{Akhmedov, Kopp, and
  Lindner}}]{Akhmedov:2008jn}
\bibinfo{author}{\bibfnamefont{E.~K.} \bibnamefont{Akhmedov}},
  \bibinfo{author}{\bibfnamefont{J.}~\bibnamefont{Kopp}}, \bibnamefont{and}
  \bibinfo{author}{\bibfnamefont{M.}~\bibnamefont{Lindner}},
  \bibinfo{journal}{JHEP} \textbf{\bibinfo{volume}{05}}, \bibinfo{pages}{005}
  (\bibinfo{year}{2008}), \eprint{0802.2513}.

\bibitem[{\citenamefont{Potzel}(2006)}]{Potzel:2006ad}
\bibinfo{author}{\bibfnamefont{W.}~\bibnamefont{Potzel}},
  \bibinfo{journal}{Phys. Scripta} \textbf{\bibinfo{volume}{T127}},
  \bibinfo{pages}{85} (\bibinfo{year}{2006}).

\bibitem[{\citenamefont{Giunti and Kim}(1998)}]{Giunti:1997wq}
\bibinfo{author}{\bibfnamefont{C.}~\bibnamefont{Giunti}} \bibnamefont{and}
  \bibinfo{author}{\bibfnamefont{C.~W.} \bibnamefont{Kim}},
  \bibinfo{journal}{Phys. Rev.} \textbf{\bibinfo{volume}{D58}},
  \bibinfo{pages}{017301} (\bibinfo{year}{1998}), \eprint{hep-ph/9711363}.

\bibitem[{\citenamefont{Bilenky
  et~al.}(2008{\natexlab{b}})\citenamefont{Bilenky, von Feilitzsch, and
  Potzel}}]{Bilenky:2008dk}
\bibinfo{author}{\bibfnamefont{S.~M.} \bibnamefont{Bilenky}},
  \bibinfo{author}{\bibfnamefont{F.}~\bibnamefont{von Feilitzsch}},
  \bibnamefont{and} \bibinfo{author}{\bibfnamefont{W.}~\bibnamefont{Potzel}}
  (\bibinfo{year}{2008}{\natexlab{b}}), \eprint{arXiv:0804.3409 [hep-ph]}.

\end{thebibliography}
\end{document}